\documentstyle[12pt]{article}
\advance \textheight by \topskip
\begin{document}

\begin{titlepage}
\hbox to \hsize{\hfil }
\hbox to \hsize{\hfil {\bf hep-ph/9611403}}
\hbox to \hsize{\hfil November, 1996}
\vfill
\large
\begin{center}
ISSUES OF SPIN PHYSICS AT RHIC
\end{center}
\vskip 1cm
\normalsize
\begin{center}
{\bf S. M. Troshin and N. E. Tyurin}\\[1ex]
{\small\it Institute for High Energy Physics,\\
  Protvino, Moscow Region, 142284 Russia}
\end{center}
\vskip 2.cm
\begin{abstract}
Some novel aspects of spin studies at RHIC are summarized
along with the persistent problems. Among them are those which
emphasize the role of angular orbital momentum in the spin
structure of the constituent quarks.
\end{abstract}
\vfill
\end{titlepage}

{\bf Introduction.} The systematic spin studies program at RHIC
have several well defined goals and among them:
\begin{itemize}
\item
study the spin structure of the nucleon, i.e., how the proton's spin
state can be obtained from a superposition of Fock states with different
numbers of constituents with nonzero spin;
\item
study how the dynamics of constituent interactions depends on  spin
degrees of freedom;
\item
understand chiral symmetry breaking and helicity non-conservation on the
quark and hadron levels;
\item
study the overall nucleon structure and long range dynamics.
\end{itemize}

 These  goals are closely interrelated in the hadron sector and
the experiments are to be interpreted in terms of hadron  spin
structure convoluted with the constituent interaction dynamics.

 The analysis of DIS data in the framework
of perturbative QCD provides information on longitudinal polarized parton
densities $\Delta q(x)$.

The study of the transverse spin structure of the nucleon
$\delta q(x)$
is equally actual  problem.
In deep inelastic  scattering the transverse
spin densities contribute only as higher--twist terms.

In hadronic interactions there could be significant spin
effects in soft processes where there is no helicity conservation rule
because the chiral $SU(3)_L\times SU(3)_R$
symmetry of the QCD Lagrangian is spontaneously broken at small $Q$.
However, the
analyzing power at low transverse momentum is small
and decreases with energy.
On the other hand, the analyzing power increases at
high transverse momentum; but this is just where we should have
$A=0$ because of the helicity conservation due to the chiral invariance
of perturbative QCD. New ideas and new experimental data are both
needed to understand the dynamics of these unexpected one-spin effects.

\bf Single-spin asymmetries in inclusive processes. \rm The study of
 spin effects in
inclusive processes probe the incoherent dynamics of hadronic
interactions. The hard production process is described in
perturbative QCD as a convolution integral of parton
cross-sections with the light--cone parton densities.
The study of spin effects in inclusive processes will yield
information on the contribution of valence, sea quarks and gluons
to the spin of the hadron:
\begin{equation}
1/2=1/2\Delta\Sigma+L_q+\Gamma+L_g
\end{equation}
In the above sum  all the terms have clear physical interpretation,
however  besides the first one, they are gauge
 and frame dependent. Transparent discussion of the theoretical aspects
of this  sum rule and a new gauge independent one are given in
 \cite{jin}.

 The primary goal of spin measurements with hadronic final states
 should be a study of onset of perturbative QCD
regime. In the spin measurements in inclusive process
$A+B\to C+X$ with polarized hadron $A$
this is based on the higher--twist origin of one--spin
transverse asymmetries \cite{kane,efremt}.
The contribution of higher twists should be small at high energies
where interactions at small distances $O(1/Q)$ can be studied.
There are some indications that such contributions are small even at
not too high energies and $Q^2$ values.
In particular, it follows
from the recent E143 data obtained at SLAC. If it is the case, the observed
significant one-spin asymmetries in hadronic processes are to be
associated with the
manifestation of nonperturbative dynamics.
However, the available
energies are not high enough to make the unambigious conclusion.
Therefore, the measurements of one-spin asymmetries at RHIC
energies are crucial. For the production of hadrons with high
$p_\perp$
        the simple
factorization is valid for the transversely polarized hadrons as well
as for the longitudinally polarized ones \cite{coll}, i.e.
\begin{equation}
A_N \sim \sum_{ab\rightarrow cd}\int d\xi_Ad\xi_B\frac{dz}{z}\delta
f_{a/A}(\xi_A) f_{b/B}(\xi_B)\hat a_N \hat \sigma
_{ab\rightarrow cd} D_{C/c}(z) \end{equation} then we should  expect
vanishing one--spin transverse asymmetries \[ A_N=0.  \]
since in the leading twist $\hat a_N= 0$.

It can also be seen expressing $A_N$ through helicity
amplitudes. Indeed,  asymmetry $A_N$ results from the interference
between helicity amplitudes which differ by one unit of helicity
\begin{equation}
A_N=2\frac{\sum_{X,\lambda _X, \lambda _2}\int d\Gamma _X
\mbox{Im}[F_{\lambda _X;+,\lambda _2}F^*_{\lambda_X;-,\lambda _2}]}
{\sum_{X,\lambda _X;\lambda _1 \lambda _2}\int d\Gamma _X
|F_{\lambda_X;\lambda _1 ,\lambda _2}|^2},
\end{equation}
where $F_{\lambda _i}$ are the helicity amplitudes of exclusive
processes. When the helicity conservation in QCD for exclusive
processes is valid at hadron level \cite{brlp}
\begin{equation}
\lambda_1+\lambda_2=\lambda_X
\end{equation}
 it follows that $A_N=0$
 in the phase  when chiral symmetry is not broken,
 i.e. in the limit of high $p_{\perp}$'s.

Possible  sources of the one-spin asymmetries are:  higher
twist effects \cite{efremt}, correlation of $k_\perp$ and spin in
structure \cite{sivs} and fragmentation \cite{cols,artru} functions,
rotation of valence quarks inside a hadron \cite{boros} and the
coherent rotation of the quark matter inside the constituent
quarks \cite{asy}.

The main points of the mechanism proposed in \cite{asy} are:
\begin{itemize} \item asymmetry reflects the internal spin structure
of the constituent quarks and is proportional to the orbital angular
momentum of current quarks inside the constituent quark; \item sign
of asymmetry and its value are proportional to polarization of the
 constituent quark inside the polarized initial hadron.
\end{itemize}

The model predicts significant one-spin asymmetries at high
$p_{\perp}$ values.
The significant asymmetries appear to show up beyond
$p_{\perp}>\Lambda _\chi(\simeq 1-2$ GeV/c), i.e.
the scale where internal structure of a constituent
quark can be probed (Fig. 1).
The observed $p_{\perp}$-behavior of
asymmetries in inclusive processes  confirms these predictions
\cite{brav}.
Thus, the study of $p_{\perp}$--dependence of one--spin
asymmetries might be used as a way to reveal the transition from the
non--pertur\-ba\-tive phase ($A_N\neq 0$) to the perturbative
phase ($A_N=0$) of QCD. The very existence of such transition can
 not be taken
for granted since the vacuum, even at
short distances, could be filled with fluctuations of gluon or quark
fields \cite{prep}. The measurements of one--spin transverse
asymmetries in this case will be important probe of the
 chiral structure of the effective QCD Lagrangian.

The relevant processes for the study of above problems are the
following:
\begin{equation}
p_{\uparrow}+{p}\rightarrow \pi^{} ,\gamma ,\mbox{jet}+X
\end{equation}

As an example we coinsider  charged pion production.
The predicted values of $A_N$  are different for the
different mechanisms. In particular, the model \cite{asy}
predicts the
energy independent values about 30\% at $p_\perp>\Lambda_\chi
(\simeq 1-2$ GeV/c), i.e.
\[
A_N^{\pi^\pm}\sim \mbox{const.} p_\perp^0\simeq \pm 30 \%
\]
at high $p_\perp$.
Note that the corresponding asymmetry in the neutral pion
production has small values due to isotopic relations.
The similar values of asymmetry predicts the model \cite{boros},
but the $p_\perp$-dependence of asymmetry should be
significantly different; however, it is not specified in the
model.
As it has been noted higher twists effects at RHIC energies and
high $p_\perp$ will not provide significant contribution to asymmetry
$A_N$. Phenomenology of these effects has not been developed yet
and predictions could vary from a few  to tens percents. However,
higher twists provide decreasing asymmetries at high $p_\perp$
\[
A_N^{\pi^\pm}\sim M/p_\perp
\]
 and could
be easily recognized then. The same is true for the predictions
of the mechanisms discussed in \cite{sivs}.

Fragmentation mechanisms
requare measurements of the two-particle final states and are relevant
for studies of transverse polarization of quarks inside polarized hadron.
This problem will be discussed later.

Experimental error in asymmetry $A_N$ is given by the events
number $N$ and the beam polarization $P_1$:
\begin{equation}
\delta A =\frac{1}{P_1}\frac{1}{\sqrt{N}}.
\end{equation}
For the values of RHIC luminosity $L=2\cdot 10^{32}$
 $cm^{-2}s^{-1}$ and beam polarization $P_1=70\%$
experimental measurements of single-spin asymmetry in pion
production with accuracy of a few percents are feasiable up
to $p_\perp$ about 30 GeV/c \cite{tan}.
It would be important to study the processes of the charged
pion production since the available data at $p_L=200$ GeV/c
indicated that the observed symmetries could be of order
of tens percents.
In the $\pi^0$ production asymmetries are expected to be smaller
due to the isospin relation between cross sections
valid in the parton model.
The $p_L=200$ GeV/c is the maximal energy
the asymmetries in these processes has been measured so far.
The study of this reaction at RHIC energies is important as a clear
test of perturbative QCD regime and nonperturbative dynamical models.
In general, these studies are important for the understanding
of the QCD vaccuum and transitions between the perturbative
and nonperturbative ones.

\bf Transverse spin densities and two--spin correlations
in inclusive processes. \rm Extensive studies of the theoretical
aspects of transverse spin
structure of the nucleon were made in  \cite{rlsp} and \cite{arme,jfji,trsv}.

The transverse quark spin density measures the difference of
 the quark momentum
distributions in a transversely polarized nucleon when a quark is
polarized parallel or antiparallel to the nucleon. This quantity
$\delta q(x)$, unlike the longitudinal quark spin
density $\Delta q(x)$,
cannot be measured in deep inelastic scattering due to its different
properties under chiral transformations.

Quark transversity is a new observable for understanding
the hadron
wave function in terms of bare quarks. Gluons give no contribution
to the transverse spin of the proton. It is promising to explore this
new spin observable and compare it with the longitudinal spin densities.

Recently a positive
 bound for $\delta q(x)$  has been
obtained \cite{sof}:
\begin{equation}
|\delta q(x)|\leq q_+(x)
\end{equation}
and bound for $\delta q(x)$ limiting its behavior at $x\to 0$ \cite{tro}:
\begin{equation}
\delta q(x)\leq\log{x}/x.
\end{equation}

The Drell--Yan process with transversely polarized protons in the
initial state
\[
p_{\uparrow}+p_{\uparrow}\rightarrow \mu ^+ + \mu ^- + X
\]
is
most suitable for the determination of the quark transversity.

The corresponding two--spin asymmetry is directly
 related to the quark
transversity distributions $\delta q(x)$.

On the other hand, we would like to stress
that the measurements
of two--spin longitudinal
asymmetries will probe the gluon contribution $\Delta g(x)$
 to the helicity of
the nucleon. The relevant processes for that purpose are the direct
$\gamma $, jets, $\chi _2$ and pion--production at high
$p_\perp$'s in the collisions of longitudinally
polarized protons.

As it has been mention transversity for spin-1/2 nucleon gets
no contribution from gluons contrary to helicity \cite{arme}.
It leads to the following signal of the onset of perturbative
QCD regime \cite{sait}
\begin{equation}
A_{TT}/A_{LL}\ll 1
\end{equation}
in the processes
\begin{eqnarray}
p+p & \rightarrow & 2 jets + X\nonumber\\
p+p & \rightarrow & \pi + X\nonumber\\
p+p & \rightarrow & Q\bar Q + X,
\end{eqnarray}
but in the Drell-Yan processes of lepton pairs production the
ratio $A_{TT}/A_{LL}$ is sensible to the ratios $\delta q/\Delta q$ and
therefore
\begin{equation}
A_{TT}/A_{LL}\sim 1.
\end{equation}

It has been stated \cite{sait}
 that the above results reflect the fundamental
aspects of parton approach in QCD: helicity, twist and
chirality selection rules.

Experimental sensitivity $\delta A_{LL/TT}$ can be estimated according
to the following formula:
\begin{equation}
\delta A_{LL/TT}=\frac{1}{P_1P_2}\frac{1}{\sqrt{N}}.
\end{equation}
As in the case of single-spin asymmetry the measurements
with accuracy of a few percents are feasiable at RHIC up
to $p_\perp$ about 30 GeV/c \cite{tan}.

\bf Strangeness in the hadrons. \rm It is  evident from deep--inelastic
scattering data that
strange quarks as well as gluons could play essential role in
the spin structure of nucleon.
 DIS data shows that strange quarks are
 negatively polarized in polarized
nucleon, $\Delta s\simeq -0.1$.
Elastic $\nu p$-scattering also indicates that strange quarks
are polarized and gives the value
 $\Delta s=-0.15\pm 0.08$ \cite{nu}.
The presence and polarization of strange quarks inside a hadron should
provide experimental signal in hadronic reactions.

We
 give first an estimate for asymmetry in
 the  production of
 $\varphi$-meson consisting of strange quarks.
It is worth to stress that in addition to $u$
 and $d$ quarks constituent quark ($U$, for example) contains pairs
of strange quarks  and the ratio of scalar
density matrix elements \begin{equation} y=
{\langle U| \bar ss|U\rangle}
/ {\langle U|\bar u u+\bar d d+\bar s s|U\rangle}
 \label{str} \end{equation} can be estimated from different
approaches  as $y=0.1 - 0.5$.

 It was argued \cite{str} that the single spin  asymmetry $A_N$ in the
process $pp\to \varphi +X$
 is connected with orbital momenta of
 strange quarks in the internal structure of constituent quarks.
The estimate for asymmetry $A_N$ in $\varphi$-meson
production at $p_\perp>\Lambda_\chi(\simeq 1-2$ GeV/c) is:
\begin{equation}
A_N(\varphi)\propto \langle{\cal{P}}_{ Q}\rangle \langle L_{\{\bar q
q\}}\rangle y\simeq 0.01-0.05.
  \label{an} \end{equation}
Thus,  a noticeable  one-spin asymmetry at high $p_{\perp}$
values in inclusive $\varphi$-meson production can be expected.
The estimate also shows that it
 is reasonable to make experimental measurements of cross-section and
asymmetry in inclusive $\varphi$-meson production to study strange content of
constituent quark as a possible source of OZI-rule evasion.

Another promising way to study the strange quark polarization inside
a nucleon is in the measurement of the hyperon polarization in the
polarized beam fragmentation region.
A very significant polarization of
$\Lambda$--hyperons has been discovered two decades ago \cite{helr}.
Since then  measurements in different processes
were performed  and number of models was
proposed for qualitative and quantitative description of these  data
\cite{xxx}. Among them the Lund model based on classical string mechanism
of strange quark pair production \cite{and}, models based on spin--orbital
interaction \cite{miet} and multiple scattering of massive strange sea
quarks
in effective external field \cite{szw} and also models for polarization of
$\Lambda$ in diffractive processes with account for proton states with
additional $\bar{s} s$ pairs such as $|uud\bar{s}s\rangle$ \cite{trt,khar}.

As it is widely known now, only part (less
than one third in fact) of the proton spin is due to quark spins
\cite{ellis,altar}.  These results can be interpreted in the
effective QCD approach ascribing a substantial part of hadron spin
to an orbital angular momentum of quark matter.
This orbital angular momentum might be revealed in
asymmetries in hadron production. The explicit mechanism has been
discussed in \cite{lamb}.
The main role  belongs to the orbital angular
momentum of $\bar q q$--pairs inside the constituent quark while
constituent quarks themselves have very slow (if at all) orbital
motion and may be described approximately by $S$-state of the
 hadron wave function.  The observed $p_{\perp}$--dependence of
$\Lambda$--hyperon polarization  in inclusive processes seems
to confirm such conclusions, since
it  appears to show up beyond
 $p_\perp>\Lambda_\chi(\simeq 1-2$ GeV/c)
 i.e.  the scale where internal structure of  constituent
quark can be probed (Fig. 2).
    The main results of the considered model:
\begin{itemize} \item polarization of
$\Lambda$ -- hyperons arises as a result of the  internal structure of the
constituent quark and its multiple scattering in the mean field. It is
proportional to the orbital angular momentum of strange quarks initially
confined in the
constituent quark;
\item sign of polarization and its value are proportional to
polarization of the constituent quark gained due to the multiple
scattering in the mean field.
\end{itemize}

It is predicted that the double spin correlation parameters should
have a similar $p_\perp$-dependence:
\[
D_{TT}\sim D_{LL}\sim 0
\]
at
 $p_\perp<\Lambda_\chi(\simeq 1-2$ GeV/c)
and
\[
D_{TT}\sim D_{LL}\sim \mbox{const.}p_\perp^0
\]
at
 $p_\perp>\Lambda_\chi $
in the polarized beam fragmentation region in the processes
\[
p_\uparrow+p\to \Lambda_\uparrow + X
\]
and
\[
p_\rightarrow+p\to \Lambda_\rightarrow + X.
\]
These relations reflects the fact that the polarized strange quark
is located
inside the constituent quark of a small size.

It is the generic feature of this model: spin effects in inclusive
processes are related to
the internal structure of constituent quark. This fact could explain
similarity in the behavior of different spin observables in inclusive
processes, i.e. rise with $p_\perp$ at small and medium transverse
momenta and then flattening at higher values of $p_\perp$.

It would be interesting to check these predictions at RHIC
energies as well as to measure for the first time triple spin
correlation parameters in the processes of hyperon production
with two polarized proton beams. It would help to understand the
mechanism of hyperon polarization.

\bf Elastic scattering. \rm In perturbative QCD, there are several
mechanisms that could give important contributions to
fixed-angle elastic scattering. However, due to small cross--section
studies of this exclusive process could be carried out in the
region where nonperturbative effects are essential.

There are  several models based on nonperturbative dynamics
predicting
significant nonzero analyzing power at fixed angles.

We quote  the
predictions of the already mentioned $U$--matrix chiral quark model
 \cite{ttel}.

In elastic scattering constituent quarks are scattered coherently
by effective field. Spin effects here are not associated with internal
structure of constituent quark as it is in the inclusive processes.
Elastic scattering then probes dynamic of scattering of constituent
quark as a whole in effective field and spin effects here  reflect
dynamics of constituent quark helicity flip in this effective field.

The analyzing power in $pp$ elastic scattering
 does not decrease with energy and
and has a nonzero value at $s\rightarrow \infty  $.
The analyzing power
at $\sqrt{s}= 0.5$~TeV and $-t=10$~(GeV/c)$^2$ in $p{p}$
elastic scattering is predicted to be 12\%, while $A$ is predicted to be 7\%
at $-t=5$~(GeV/c)$^2$.

Other nonperturbative models \cite{mqm,diq} also
predict nonzero values for the analyzing power in RHIC energy range, while
perturbative QCD (hard scattering model) predict vanishing values.

For estimation of the experimental sensitivity  the following dependence
of differential cross--section of $pp$--elastic scattering
provided by the mentioned model for the fixed
 $t$ region ($-t\gg 1$, $-t/s\ll 1$) can be used:
\begin{equation}
\frac{d\sigma}{dt}\simeq \frac{R(s)}{\sqrt{-t}}\exp[-\frac{2\pi\xi}{M}
\sqrt{-t}],
\end{equation}
where $M=Nm_Q$, $N=6$, $m_Q=0.3$ GeV and $\xi=1.7$,
$R(s)\sim \frac{1}{M}\ln s$ is
the interaction radius.
Such dependence is in a very good agreement with the available
experimental  at $ 3<-t<12$ (GeV/c)$^2$ and $p_L>400$ GeV/c.

The measurement of analyzing power $A$ in the elastic
scattering process
$ p_{\uparrow}+{p}\rightarrow p+{p}$
at RHIC will allow:
\begin{itemize}
\item
study of hadron structure at long distances as a bound state of the
constituent quarks and role of confinement;
\item
study of the dynamics of constituent quark helicity flip interactions
and the role of spontaneous breaking of chiral symmetry.
\end{itemize}

Estimates of experimental sensitivity show that the region of
$-t\sim 10$ (GeV/c)$^2$ is experimentally accessible.

\bf Conclusion. \rm Spin measurements at RHIC with  polarized proton beams
would probe the fundamental couplings of the underlying Lagrangian
and investigate the spin structure of the nucleon.
 A wide range of one-- and two--spin asymmetries could be
measured. As has often happened in the past,
these spin measurements might bring unexpected new results;
this would certainly stimulate the development of new
theoretical ideas.

\vspace{0.5cm}
These notes have been prepared in the framework of
the PHENIX/Spin experiment.
We are pleased to thank Mike Tannenbaum for his helpful comments and
reading the manuscript.

\newpage \normalsize \begin{center} \bf Figure captions \end{center}
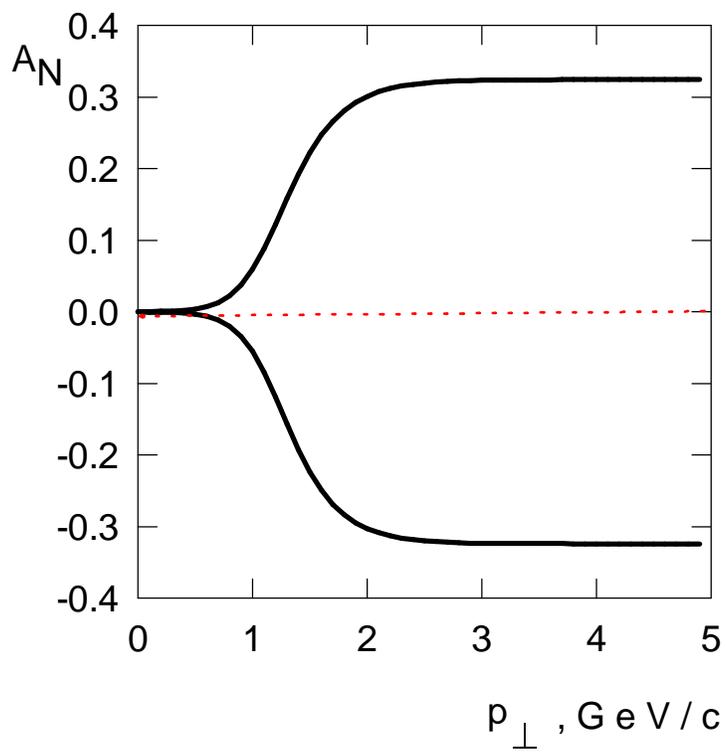
\rm \bf Fig. 1. \rm Asymmetry $A_N$ in the process
$p_{\uparrow}+p\rightarrow \pi^++X$ (positive values) and in the
process $p_{\uparrow}+p\rightarrow \pi^-+X$ (negative values) at
$\sqrt{s}=500$ GeV.\\[2ex]
\bf Fig. 2 \rm The $p_\perp$--dependence of $\Lambda$--hyperon polarization
in the process $pp\rightarrow\Lambda X$ at $p_L=400$ GeV/c.
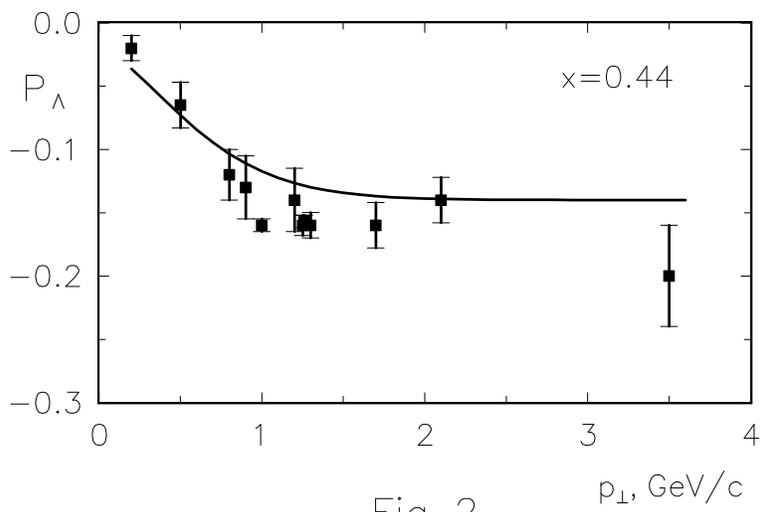
\end{document}